\documentstyle[12pt]{article}
\begin{document}
{\noindent Mod. Phys. Lett. {\bf A 16} (2001) 789--794}
\vspace{12mm}
\begin{center}
{\large \bf Gravitational Effects of Rotating Bodies}
\footnote{The project supported partially by the Ministry of
Science and Technology of China under Grant Nos. 95-Yu-34 and
NKBRSF G 19990754 and the National Natural Science Foundation of
China under Grant Nos. 10047004 and 19835040} \\

\vspace{8mm}
Yuan-Zhong Zhang$^{a}$, Jun Luo$^b$, and Yu-Xin Nie$^c$\\
\vspace{4mm}
   {\footnotesize{\it
 $^a$Institute of Theoretical Physics, Chinese Academy of Sciences,
           P.O. Box 2735, Beijing, China\footnote{
            Email: yzhang@itp.ac.cn}\\
 $^b$Department of Physics, Huazhong University of Science and Technology,
          Wuhan, China \\
 $^c$Institute of Physics, Chinese Academy of Sciences,
            Beijing, China}}\\

\end{center}

\vspace{8mm}

\begin{abstract}
We study two type effects of gravitational field on mechanical gyroscopes
(i.e. rotating extended bodies). The first depends on special relativity and
equivalence principle. The second is related to the coupling (i.e. a new force)
between the spins of mechanical gyroscopes, which would violate the equivalent
principle. In order to give a theoretical prediction to the second we suggest
a spin-spin coupling model for two mechanical gyroscopes. An upper limit on the
coupling strength is then determined by using the observed perihelion precession
of the planet's orbits in solar system. We also give predictions violating the
equivalence principle for free-fall gyroscopes .
\end{abstract}

\vspace{10mm}

It is known that the gravitational couplings of intrinsic spins of elementary
particles have been investigated for long time in both theory and experiment
(see, e.g., Ref. [1] and the literature quoted there). On the other hand,
some experiments have been performed by use of mechanical gyroscopes and give
contradictory results [2-6]. However, in theory the spin of a mechanical gyroscope
(i.e. the rotation of an extended body) is essentially different from the intrinsic
spin of an elementary particle. A reasonable deduction is then that the gravitational
couplings of intrinsic spin could not be used to describe those of the spin of
mechanical gyroscope. Thus it is necessary to study theoretically such effects of
possible spin-gravitational couplings on the motion of the center-of-mass for
a mechanical gyroscope in an external gravitational field. This is just the aim of
the present paper.

For a rotating body (a mechanical gyroscope) in an external gravitational field
there might exist two different kinds of physical effects. The first is that its
inertial mass, according to the mass-velocity relation in special relativity,
would increase owing to its rotation. And correspondently there would also be an
increment in its gravitational mass. Hence a new test of the mass-velocity relation
and equivalence principle could be given by comparing the gravitational acceleration
of center-of-mass of a rotating extended body (i.e. a mechanical gyroscope) with that
of an irrotating body moving in an external gravitational field. In addition, there
would exist another sort of physical effect between mechanical gyroscopes, i.e.
a new force between two rotating bodies, which would violate the equivalence principle.

We now discuss the first kind of effect. Consider a rotating rigid ball. According
to special relativity the inertial mass of the rotating rigid ball is given by
  $$m^I=\int  {\rm d^3}x\frac{\rho}{\sqrt{1-v^2 /c^2}}\simeq
     m+\frac{m}{5}\left(\frac{\Omega R}{c}\right)^2,      \eqno(1)$$
where the mass density $\rho$ is assumed to be a constant, $\Omega$ and
$R$ are, respectively, angular velocity and radius of the rotating ball and
$\Omega R \ll c$, $m$ is the inertial mass of the ball when it was not rotated
(we call it the rest mass), and $c$ is the speed of light.

In order to test the prediction of special relativity, we introduce the
parameter $\eta^{sp}$ in the second term of the right-hand side of Eq. (1),
i.e.,
  $$m^I = m\left[1+\eta^{sp}\frac{1}{5}\left(\frac{\Omega R}{c}\right)^2
           \right].                                          \eqno(2)$$

The passive gravitational mass of the rotating ball is assumed to be
  $$M^p = m\left[1+\eta^{ep}\eta^{sp}\frac{1}{5}\left(\frac{\Omega
            R}{c}\right)^2\right],                          \eqno(3)$$
where $\eta^{ep}$ is a dimensionless parameter, of which a deviation from unity
indicates a violation of the equivalence principle. Here validity of the equivalence
principle for irrotational bodies is proposed.

Then the free-fall acceleration of the rotating body within an external gravitational
field could be written as
  $$a=g\frac{M^p}{m^I}\simeq g\left[1+(\eta^{ep}-1)\eta^{sp}
   \frac{1}{5}\left(\frac{\Omega R}{c}\right)^2 \right].    \eqno(4)$$
where $g$ is gravitational acceleration of the external gravitational field.
Comparing the rotating ball to another irrotational ball with the
same rest mass $m$, we have the relative free-fall acceleration
between them,
  $$\frac{\Delta a}{a}\simeq\frac{(\eta^{ep}-1)\eta^{sp}}{5}
   \left(\frac{2\pi Rn}{c}\right)^2.                        \eqno(5)$$
For $n=4\times 10^4 {\rm rpm}, R=10{\rm cm}$ one arrives at from Eq. (5)
   $$\frac{\Delta a}{a}=(\eta^{ep}-1)\eta^{sp}\times 4\times 10^{-13}.
                                                              \eqno(6)$$

It is shown from Eq. (6) that $(\eta^{ep}-1)\eta^{sp}\not= 0$
implies violation of the equivalence principle for the rotating body.
Thus comparing accelerations of a mechanical gyroscope and an irrotating body free
falling in Earth's gravitational field could give a test of the equivalence principle
together with the mass-velocity relation. It is note that from this sort of observation
one could not distinguish the test of the equivalence principle from that of the
mass-velocity relation.

We next study the second kind of effect, the spin-gravitational coupling between two
mechanical gyroscopes, which would violate the equivalence principle. It is known that
Newtonian gravitational theory defines spin angular momentum for a extended body, but
does not give spin-gravitational coupling. General relativity predicts spin-spin couplings
between extended bodies, which causes spin precession but does not violate the equivalence
principle for the motion of the center-of-mass. In addition there are many other
gravitational theories, such as Poincare gauge theory of gravitation, high-dimension
gravitational theories, superstring theory, and etc, all of which deal with the
spin-gravitational couplings of intrinsic spins of particle fields. Because the intrinsic
spins essentially differ from the rotation of mechanical gyroscopes, so it is difficult to
discuss the interaction between the mechanical gyroscopes within the framework of these
theories. In this paper we shall develop a phenomenological model for the interaction
between the rotating bodies, which is independent of any specific theory.

We assume that spin-interaction exists between two rotating rigid bodies.
Owing to the spin-coupling the rotating bodies would get an extra-energy
(an extra-potential) $E^s$, and then have the extra inertial mass from
the mass-energy relation in special relativity:
  $$m^s = E^s /c^2.                                     \eqno(7)$$
The rotating body might then get the following raise in its passive
gravitational mass:
  $$M^s = \eta^s E^s /c^2,                              \eqno(8)$$
where $\eta^{s}$ is a dimensionless parameter.
The total inertial masses $m^I_i$ and the total passive gravitational
masses $M^p_i$ of the rotating bodies are then
  $$m^{I}_{i}= m_{i} +  E^s_{i} /c^2,                    \eqno(9)$$
and
  $$M^p_{i} = m_{i} + \eta^s E^s_{i} /c^2,               \eqno(10)$$
respectively, where $m_{i}$ for $i=1,2$ stand for the rest-masses of
the bodies when they were not rotated. Here we have assumed that
the equivalence principle is available for irrotational bodies.
The equations of motion for the rotating bodies are then
 $$m^I_{i}{\bf a}_{i}=-\frac{GM^{p}_{i}M^{p}_{j}}{r_{ij}^3}{\bf r}_{ij},
                                                        \eqno(11)$$
where ${\bf r}_{ij}={\bf r}_{i}-{\bf r}_{j}$.
Using Eqs. (9) and (10) and considering $E^s_{i}/m_{i}c^2 \ll 1$, the
equation (11) can be written approximately as
 $$m_{i}{\bf a}_{i}=-\frac{Gm_{i}m_{j}}{r_{ij}^3}{\bf r}_{ij}
     \left[1+(2\eta^s -1)\frac{E^s_i}{m_i c^2}\right].
                                                        \eqno(12)$$
In order to satisfy the motion theorem for the center of mass of
the two body system and the theorem of moment of momentum, we have
assumed ${E^s_1}/(m_1 c^2)={E^s_2}/(m_2 c^2)$.

The extra-energy $E^s$ generated by the interaction between the rotating
bodies has not yet been predicted in gravity theories as mentioned above.
We here give a phenomenological model for the spin-gravitational interaction of
rotating bodies. It is natural to suppose the extra-energy $E^{s}$ being proportional
directly the gravitational potential $V(1/r)$. In other words the extra-energy is
a function of $1/r$, which is assumed to be the form
$$E^{s} = E^{s}\left( \frac{1}{r}\right)=\alpha \frac{1}{r} +\beta \frac{1}{r^2}+  \cdots $$
where $\alpha$ and $\beta$ depend on spin-angular momentium of the rotating bodies.
For simplicity we here consider only the first term on the right-hand side of
the above equation, and then write
  $$\frac{E^s_1}{m_1 c^2}=\frac{E^s_2}{m_2 c^2}={\rm g}_{s}\frac
     {{\bf s}_1 \cdot {\bf s}_2}{Gm_1 m_2}\frac{1}{r_{12}},\eqno(13)$$
where ${\bf s}_1 $ and ${\bf s}_2 $ are, respectively, the spin angular
momentum of the rotating bodies 1 and 2, and ${\rm g}_s $ is the coupling
constant for the spin interaction.

By using Eq. (13) in Eq. (12), we have the gravitational force between the
center-of-mass of two mechanical gyroscopes as follows:
  $$m_{i}{\bf a}_{i}=-\frac{Gm_{1}m_{2}}{r_{12}^3}{\bf r}_{ij}
     -(2\eta^s -1){\rm g}_{s}\frac{{\bf s}_1 \cdot {\bf s}_2}
       {r_{12}^4}{\bf r}_{ij},                             \eqno(14)$$
for $i,j=1,2$. The second term on the right-hand side of the equation is a new force,
i.e. the spin-gravitational coupling of the gyroscopes, which would give perihelion
precession of the planet's orbits in solar system and also violate the equivalence
principle. Thus the observations of the perihelion precession could give an upper
limit on the parameter $(2\eta^s -1){\rm g}_{s}$. For this purpose we apply
Eq. (14) for the planet's orbit in solar system, and get the following solution
to Eq. (14)
  $$r=\frac{(1-\delta)p}{1+e {\rm cos}(\sqrt{1-\delta}\phi -\omega)},
                                                           \eqno(15a)$$
  $$p=\frac{C^2}{G\left(M_{\odot} +M\right)}=\frac{1}{2}
        \left(r_{min}+r_{max}\right)(1-e^2)+O(\delta),     \eqno(15b)$$
and
  $$\delta =(2\eta^s -1){\rm g}_{s}\frac{{\bf s}_\odot \cdot {\bf s}}
           {GM_{\odot}M}\frac{1}{p},                      \eqno(15c)$$
where $M_{\odot}$ and $M$ are the masses of, respectively, the Sun and
planet, $e$ is the eccentricity of the planet's orbit,
$G$ is Newtonian gravitational constant,
and $C$ is an integration constant.

We see that non-vanishing $\delta$ would give perihelion precession.
In fact, for the perihelion we have ${\rm cos}(\sqrt{1-\delta}\phi-
\omega )=1$, i.e., $\sqrt{1-\delta}\phi_n -\omega = 2n\pi$ with
$n=0,1,2, \cdots$. We then get the precession:
    $$\Delta \phi =\phi_n -\phi_0 -2n\pi =2n\pi \left(\frac{1}
          {\sqrt{1-\delta}}-1\right)\simeq n\pi\delta.  \eqno(16)$$
Using the observations of the planet's perihelion precessions, in particular the
observed Earth's perihelion precession $(5.0\pm 1.2)$ arcseconds
per one hundred years, we obtain a limit on the theoretical result of
Eq. (16): $100\pi\delta\leq 1.2$. Explicitly we have
  $$(2\eta^s -1){\rm g}_{s}{\rm cos}\chi\leq 3.4\times 10^{-31}
      {\rm gram}^{-1},                                     \eqno(17)$$
where $\chi$ is the angle between the spin directions of the Sun and
Earth.

Let us now consider two rigid balls falling freely in Earth's
gravitational field, one of them rotating and the other
irrotational.  Relative difference in accelerations of the two
balls can be obtained from Eq. (14)
  $$\frac{\Delta a}{a}\equiv \frac{a_2 -a_1 }{a_2}\simeq
    (2\eta^s -1){\rm g}_{s}\frac{{\bf s} \cdot {\bf s}_\oplus}
    {Gm M_\oplus R_\oplus},                                 \eqno(18)$$
where $m$ and ${\bf s}$ are the mass and angular momentum of the rotating
ball, $M_\oplus, R_\oplus$ and ${\bf s_\oplus}$ are the mass, radius and
angular momentum of the Earth, and the parameter $(2\eta^s -1){\rm g}_{s}$
is given by Eq. (17). For a rotating rigid ball with constant mass density,
we have $|{\bf s}|=(4/5)\pi nR^2 m$, where $R$ and $n$ are the radius
and speed of rotation. For $n=4\times 10^4 {\rm rpm}$ and $R=10$cm,
one arrives at from Eq. (18)
  $$\frac{\Delta a}{a}\leq 2\times 10^{-14}.                \eqno(19)$$

It is note that the second prediction (19) follows from a different physical
source rather than the first one (6). The second effect (19) could be
distinguished experimentally from the first one in such a way: The second
depends on the relative directions of the two gyroscopes while the first
does not.

It is mentioned that in 1989 Hayasaka and Takeuchi [2] reported a balance
experiment in which the right spinning gyroscope induced a weight decrease
proportional to the rotational velocity, whereas the left spinning caused no
weight change. Later the subsequent experiments (1990) [3-5] gave contradictory
results: no weight change of gyroscopes when measured by balances, which within
uncertainty are in agreement with the predictions above. Recently Hayasaka
et al. [6] reported a free-fall experiment using a spinning gyro, in which a
similar positive result was obtained: the mean value of the fall-accelerations
of the right-spinning at 18 000 rpm is significantly smaller than that of zero
spinning. This means that the equivalence principle is not valid for a
right-spinning gyroscope. However, the weight change in the experiment is too
larger than that predicted by the effects above. Thus it is necessary to
repeat this kind of experiment.

These two effects, Eqs. (6) and (19), could be observations by use of the
rotating torsion balances, satellites or long-distance free-fall experiments.
For example, for a proposed free-fall experiment the drop test masses may be
the inner spinning sphere and outer shell consisting of a superconductor
gyroscope, and the gaps between the inner and outer test masses could be
measured using the SQUID technology. The superconductor gyroscope is suspended
in the top of a vacuum outer capsule. Make the inner sphere free-fall by
turning off the power supply of the gyroscope during the outer shell free moving
in the vacuum outer capsule after cutting off the suspension. This proposed
experiment at a drop tower of one hundred meter high could reach an accuracy of
$\Delta a/a < 10^{-12}$ [7]. A more longer-distance free-fall experiment
could be done in such a way: A balloon can lift the facilities to about
40 kilometers altitude and then drop it down. There will be about 40 seconds
free drop time for the test masses, which may improve the accuracy to
$< 10^{-13}$ sensitive to the effects predicted above.

\vspace{5mm}
\noindent
{\large \bf Acknowledgments}

\vspace{2mm}
The authors would like to thank Professors Wen-Rui Hu, Yuan-Ben Dai,
Zhong-Yuan Zhu, Tao Huang, and Zhen-Long Zou for their helpful
discussion.

\end{document}